\documentclass[aps,twocolumn]{revtex4}

\usepackage{graphicx}
\usepackage{dcolumn}
\usepackage{bm}
\usepackage{amsmath}
\begin{document}

\newcommand{\bk}{{\bf k}}
\newcommand{\bp}{{\bf p}}
\newcommand{\bv}{{\bf v}}
\newcommand{\bq}{{\bf q}}
\newcommand{\tbq}{\tilde{\bf q}}
\newcommand{\tq}{\tilde{q}}
\newcommand{\bQ}{{\bf Q}}
\newcommand{\br}{{\bf r}}
\newcommand{\bR}{{\bf R}}
\newcommand{\bB}{{\bf B}}
\newcommand{\bA}{{\bf A}}
\newcommand{\bK}{{\bf K}}
\newcommand{\vd}{{v_\Delta}}
\newcommand{\tr}{{\rm Tr}}
\newcommand{\kslash}{\not\!k}
\newcommand{\qslash}{\not\!q}
\newcommand{\pslash}{\not\!p}
\newcommand{\rslash}{\not\!r}
\newcommand{\bs}{{\bar\sigma}}

\title{Theory of the quasiparticle interference patterns in the 
pseudogap phase of the cuprate superconductors}

\author{T. Pereg-Barnea and M. Franz}
\affiliation{Department of Physics and Astronomy,
University of British Columbia, Vancouver, BC, Canada V6T 1Z1}
\date{\today}

\begin{abstract}
A method is proposed to test for the nature of the pseudogap phase in 
cuprates using the recently developed technique of Fourier transform 
scanning tunneling spectroscopy. We show that the observed quasiparticle 
interference patterns depend critically on the quasiparticle coherence 
factors, making it possible to distinguish between the pseudogap 
dominated by superconducting fluctuations 
and by various particle-hole condensates. 
\end{abstract}
\maketitle
Existence of the pseudogap phase, a non-superconducting phase with 
suppressed single particle density of states
(DOS)  \cite{timusk1}, represents perhaps the most dramatic departure 
of cuprates from the
Landau Fermi liquid BCS-Eliashberg paradigm believed to describe 
conventional low-$T_c$ superconductors. Theoretical approaches to the 
pseudogap phenomenon in  cuprates fall broadly into two classes. 
One school of though
attributes it to the incipient superconducting order whose 
amplitude forms at $T^*>T_c$ but remains phase incoherent down to
$T_c$ \cite{emery1,randeria1,fm1,balents1,levin1,ft1,ftv1,laughlin1}. 
The other school ascribes the pseudogap to the formation of
some other (static or fluctuating) order, usually in the
particle-hole (p-h)
channel \cite{so5,varma1,ddw1,vojta1}. Progress in understanding 
the physics of cuprates, and perhaps other strongly correlated 
superconductors, depends on the successful determination of the 
origin of the pseudogap phenomenon.

In this Communication we propose a test for the nature of the pseudogap
phase based on the study of the quasiparticle interference patterns  
that can be observed by means of Fourier transform scanning tunneling
spectroscopy (FT-STS) \cite{Hoffman,McElroy,Howald}. The existing 
data, taken deep in the superconducting phase of 
Bi$_2$Sr$_2$CaCu$_2$O$_{8+\delta}$ (Bi-2212), exhibit distinctive
patterns of peaks in the reciprocal space. These peaks disperse 
through the 
Brillouin zone in a manner consistent with the band structure
deduced from the angle resolved photoemission spectroscopy (ARPES). 
In what follows we demonstrate that, contrary to the existing
consensus, the patterns observed in FT-STS depend in a crucial
way on the quasiparticle {\em coherence factors}. We show that 
the appearance of peaks at the particular positions is unique to 
the superconducting order, and reflects the pairing (p-p) character
of the coherence factors. By contrast the interference patterns   
in a state with p-h ordering will be qualitatively different in that
peaks appear at different $\bk$-vectors or not at all. Our proposal
consists of extending the FT-STS measurements into the pseudogap
phase above $T_c$. Observation of patterns similar to those 
seen well below $T_c$ would then imply that pseudogap is 
predominantly of superconducting origin. Observation of qualitatively 
different patterns discussed below would imply order of another type. 
Preliminary data \cite{Yazdani} appear to 
support the former scenario.

\begin{figure}
\includegraphics[width = 7.8cm]{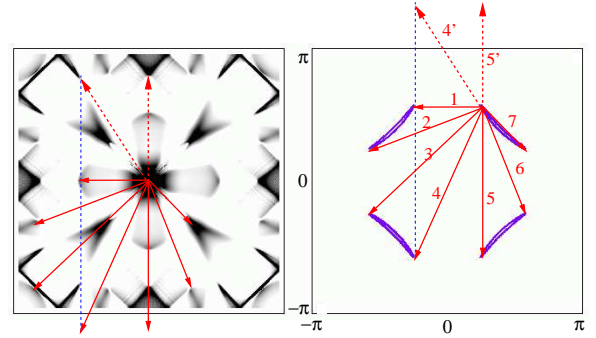}
\caption{Right: The banana shape contours of constant low energy around 
the nodes of 
the gap and seven vectors connecting their ends \cite{Hoffman}. 
Left: Numerical evaluation of the Fourier transformed local density of 
states in Eq. (\ref{eq:ldos}) with $t' = 0.3 t$, 
$\Delta_0 = 0.2t$, $\mu = -t$ and $\omega = 0.45\Delta_0$. 
In order to simulate the experimental resolution we filtered out the 
low-intensity background signal. The peaks 
disperse in a manner consistent with the 
octet model and the experimental data \cite{Hoffman,McElroy}.
}\label{fig:dSC}
\end{figure}

The existing FT-STS results \cite{Hoffman,McElroy} have been interpreted
via the `octet model' \cite{Wang}, illustrated in Fig.\ 
\ref{fig:dSC}.
This model asserts that the interference patterns arise
due to the elastic quasiparticle scattering from random disorder between 
the regions in the Brillouin zone with high DOS. In a $d$-wave 
superconductor ($d$SC) these are situated at the ends of 
the banana-shaped contours of constant energy and lead to seven 
characteristic vectors $\bq_i$. This octet model works remarkably 
well in describing 
the data and furthermore agrees with detailed numerical studies 
of the interference patterns \cite{Wang,Polkov,Capriotti}. 
Our first step will therefore be to understand how the 
interference pattern is formed in the superconducting phase. 

The local density of states (LDOS) in a superconductor  
$n(\br, \omega)$ is given by the particle-hole part 
of the Nambu Green's function. 
Within the usual $T-$matrix formulation \cite{Wang,Polkov} the Fourier 
transformed LDOS modulation $\delta n(\bq,\omega)$ can be expressed as 
\begin{equation}\label{eq:ldos}
\delta n = {\rm Im}{1\over L^2}\sum_\bk [G_0(\bk,\omega)
\hat{T}(\bk,\bk-\bq,\omega)G_0(\bk-\bq,\omega)]_{11},
\end{equation}
where $L$ is the linear size of the system and $G_0(\bk,i\omega)=
[i\omega-\epsilon_\bk\tau_3- \Delta_\bk\tau_1]^{-1}$ is the 
superconducting Green's function in the Nambu space. In what follows 
we consider a model with dispersion  
$\epsilon_\bk = -2t(\cos k_x+\cos k_y)
-4t'\cos k_x \cos k_y-\mu$ and a $d$-wave gap  
$\Delta_\bk = {\Delta_0 \over 2}(\cos k_x -\cos k_y )$.
As shown by Capriotti {\em et al.\ } \cite{Capriotti} the
interference pattern measured in FT-STS for weak non-magnetic impurities
is encoded in the quantity
\begin{equation}\label{eq:lam1}
\Lambda(\bq,\omega) = {1\over L^2}\sum_\bk [G_0(\bk,\omega)\tau_3
G_0(\bk-\bq,\omega)]_{11}.
\end{equation}
For simplicity we start by analyzing this case and we return to Eq.\
(\ref{eq:ldos}) with the full $T$-matrix shortly. 

It is instructive to evaluate 
$\Lambda(\bq,\omega)$ in the low energy limit $\omega\ll\Delta_0$, where 
analytic results can be obtained. Inserting $G_0$ into Eq.\ (\ref{eq:lam1})
we find
\begin{equation}\label{eq:lam2}
\Lambda(\bq,i\omega) = {1\over L^2}\sum_\bk 
{(i\omega+\epsilon_+)(i\omega+\epsilon_-)-\Delta_+\Delta_-\over
(\omega^2+E_+^2)(\omega^2+E_-^2)},
\end{equation}
with $\epsilon_\pm=\epsilon_{\bk\pm\bq/2}$, 
$\Delta_\pm=\Delta_{\bk\pm\bq/2}$ and 
$E_\pm=\sqrt{\epsilon_\pm^2+\Delta_\pm^2}$.
We now focus on the situation when both $\bk\pm{1\over 2}\bq$
are close to one nodal point; this should contribute
to the peak in $\Lambda(\bq,\omega)$ at the vector labeled as $\bq_7$
in the  Fig.\ \ref{fig:dSC}. Near this node we define a local
coordinate system $(k_1,k_2)$ and linearize the dispersion 
in the usual manner;
$\epsilon_\bk\to v_Fk_1$ and $\Delta_\bk\to v_\Delta k_2$, where 
$v_F$ and $v_\Delta$ are quasiparticle velocities perpendicular and
parallel to the Fermi surface, respectively. We find
\begin{equation}\label{eq:lam3}
\Lambda_{\rm lin} = {1\over v_Fv_\Delta}\int {d^2k\over(2\pi)^2} 
{-\omega^2 + (k_1^2-k_2^2)-(\tq_1^2-\tq_2^2)\over
[\omega^2+(\bk-\tbq)^2][\omega^2+(\bk+\tbq)^2]},
\end{equation}
where we have scaled the integration variables $k_1\to k_1/v_F$, 
$k_2\to k_2/v_\Delta$ and defined a new vector $\tbq={1\over 2}
(v_F q_1,v_\Delta q_2)$.

The above rescaling leads to our first useful insight: in the scaled 
frame of reference the contours of constant energy are concentric
circles implying that vector $\bq_7$ cannot have any special significance
as far as the DOS is concerned. As we shall see below the peaks at 
$\pm\bq_7$
arise solely due to the {\em coherence factors}, i.e.\ factors appearing
in the numerator of Eqs.\ (\ref{eq:lam2},\ref{eq:lam3}).

\begin{figure}
\includegraphics[width = 8cm]{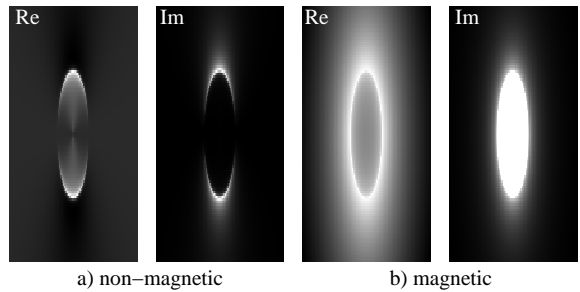}
\caption{Interference patterns in the single-node approximation. a)
non-magnetic scattering Eq.\ (\ref{eq:lam4}), b) magnetic scattering
Eq.\ (\ref{eq:lam5}). In both cases we use $v_F/v_\Delta=4$. Note that 
the axes here represent the nodal coordinates 
$q_1$ (horizontal) and $q_2$ 
(vertical), rotated by $45^\circ$ relative to Fig.\ \ref{fig:dSC}.}
\label{fig:lin}
\end{figure}

Integrals of the type appearing in
 Eq.\ (\ref{eq:lam3}) are familiar from the theory
of massless relativistic Dirac fermions \cite{peskin} and can be
evaluated most conveniently by exploiting the Feynman parameterization
\cite{note1}. The exact result is
\begin{eqnarray}\label{eq:lam4}
\Lambda_{\rm lin}(\bq,\omega) &=&{1\over 2\pi v_Fv_\Delta}
\left[\left({\tq_2\over\tq}\right)^2 
{\cal F}\left({\omega\over\tq}\right)
-{1\over 2}\right],\nonumber\\
{\cal F}(z)&=&1-{z^2\over\sqrt{z^2-1}}\arctan{1\over\sqrt{z^2-1}}.
\end{eqnarray}
For a given fixed energy $\omega$
function  ${\cal F}(z)$ implies an inverse square root singularity 
in both the real and
imaginary parts of $\Lambda_{\rm lin}(\bq,\omega)$ along an elliptic 
contour of constant energy given by $E_\bq=2\omega$ (here
$E_\bq=2\tq\equiv\sqrt{v_F^2q_1^2+v_\Delta^2q_2^2}$). More importantly
this singularity is weighted by an angular factor $(\tq_2/\tq)^2=
(v_\Delta q_2/E_\bq)^2$ producing the largest amplitude at the two ends
of the ellipse, as illustrated in Fig.\ \ref{fig:lin}. These points
of largest intensity coincide with $\pm\bq_7$. We conclude 
that in this case the octet model works because of the {\em special BCS 
coherence factors} and not because of the DOS arguments.
One can perform similar
analyses for the internodal scattering and find a large response near
some of the vectors $\bq_i$ indicated in Fig.\ \ref{fig:dSC}. Interestingly 
we find that the peak expected from the
octet model near $\bq_4$, which is absent in the 
experimental data \cite{McElroy}, corresponds to an endpoint of a line of
higher intensity. 

One may now inquire what is the physics leading to the angular 
modulation of the singularity displayed in Eq.\ (\ref{eq:lam4})
and the consequent formation of the interference peaks in 
$\Lambda(\bq,\omega)$. A little thought reveals that the principal cause
of this behavior is the particle-hole mixing. Charge is not 
a good quantum number in a superconductor; non-magnetic impurities couple
to charge and this leads to strong angular modulation of the 
quasiparticle response as one travels along the contour of constant 
energy around the nodal point. To see this more clearly consider 
scattering by {\em magnetic} impurities. Magnetic impurities couple to 
quasiparticle {\em spin} and the latter remains a good quantum number
in a superconductor. To describe this situation
we replace $\tau_3$ in Eq.\ (\ref{eq:lam1}) by $\tau_0\equiv \openone$. This
reverses the sign in front of the 
$\Delta_+\Delta_-$ term in Eq.\ (\ref{eq:lam2}). Such a sign reversal
has profound consequences for the response function $\Lambda(\bq,\omega)$
which can be again evaluated in the low energy limit near a single node.
In this case the result is 
\begin{eqnarray}\label{eq:lam5}
\Lambda_{\rm lin}^{\rm mag}(\bq,\omega) &=&{-1\over 2\pi v_Fv_\Delta}
\left[i\pi + 2 {\cal F}'\left({\omega\over\tq}\right)
+\ln\left({\omega^2\over\lambda^2}\right)\right],\nonumber \\
{\cal F}'(z)&=&\sqrt{z^2-1}\arctan{1\over\sqrt{z^2-1}},
\end{eqnarray}
with $\lambda\simeq \Delta_0$ a high energy cutoff.
The inverse square root singularity of Eq.\ (\ref{eq:lam4}) 
is replaced by a cusp 
along $E_\bq=2\omega$ and, crucially, there is now no angular 
modulation. The interference patterns for magnetic disorder 
will exhibit {\em continuous lines} near the center of the 
BZ (Fig.\ \ref{fig:lin}b) and will be qualitatively
different from those with scalar disorder, as already noted in Ref.\
\cite{Wang}.

The analysis of the low-energy limit of $\Lambda(\bq,\omega)$ thus 
shows that while the quasiparticle 
dispersion $E_\bq$ determines the possible loci of strong interference,
it is the coherence factors that determine the strength and select the 
actual location of the peaks on these loci.   

To conclude our discussion of the superconducting state we evaluate
$\delta n(\bq,\omega)$ given by Eq.\ (\ref{eq:ldos})
numerically with the full dispersion $\epsilon_\bk$ and 
 parameters relevant to Bi-2212. In Fig.\ \ref{fig:dSC} we use 
the full single-impurity, intermediate phase shift
$T$-matrix, $\hat{T}(\omega)=T_0(\omega)\tau_0+T_3(\omega)\tau_3$,
as described in Ref.\ \cite{balatsky1}. We find that experimental patterns
are best reproduced by a particular value of the phase shift that 
yields $T_3(\omega)\approx-T_0(\omega)^*$. This choice effectively kills the 
analog of the $\Delta_+\Delta_-$ term in Eq.\ (\ref{eq:lam2}) and
we are led to believe that in real systems this cancellation has a more 
fundamental cause, perhaps related to spatial fluctuations of the
phase of $\Delta$ in a material with nanoscale electronic inhomogeneity
\cite{Hoffman,McElroy}.  
\begin{figure}
\includegraphics[width = 8cm]{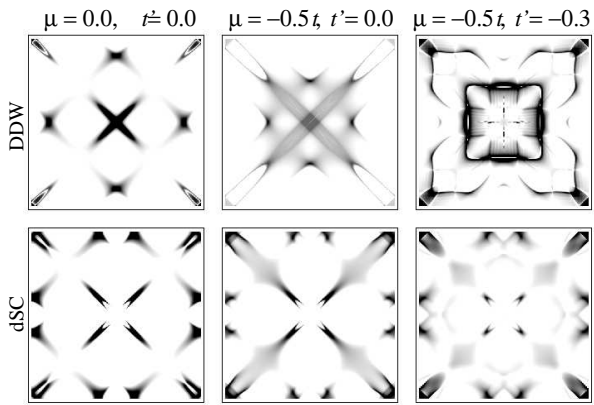}
\caption{ Interference patterns in a DDW 
model of the pseudogap (top) compared with a $d$SC (bottom). 
In all figures $\omega = 0.3\Delta_0$ and $\Delta_0=D_0 = 0.5t$. 
}
\label{fig:DDW}
\end{figure}

We now turn to the pseudogap state. If the pseudogap is due to an 
ordering in the p-h channel then
the coherence factors will generally differ from those describing
the superconductor. It is clear from the previous 
discussion that such coherence 
factors will produce qualitatively different interference patterns
even if the DOS remains similar to that of a $d$SC. We now illustrate 
this general statement on the example of a $d$-density wave (DDW)
state \cite{ddw1} that has been proposed to describe the 
pseudogap phase in cuprates. This is perhaps the most relevant example
since for specially chosen parameters $(\mu=t'=0)$ 
the DOS in DDW state is 
{\em identical} to that in $d$SC. We show below that even in this case
FT-STS patterns are qualitatively different.

Under the assumptions leading to Eq.\ (\ref{eq:lam1}) one can
show that for DDW state the interference pattern is given by
\begin{equation}\label{eq:lam6}
\Lambda(\bq,\omega) = {1\over L^2}{\sum_\bk}' 
{\rm Tr}[G_0(\bk,\omega)(1+\tau_1)G_0(\bk-\bq,\omega)],
\end{equation}
where the prime denotes summation over the reduced Brillouin zone. The 
DDW propagator
reads $G_0(\bk,i\omega)=[(i\omega-\epsilon'_\bk)-\epsilon''_\bk
\tau_3 -D_\bk\tau_2]^{-1}$, with 
$\epsilon'_\bk={1\over 2}(\epsilon_\bk+\epsilon_{\bk+\bQ})$,
$\epsilon''_\bk={1\over 2}(\epsilon_\bk-\epsilon_{\bk+\bQ})$,
$\bQ=(\pi,\pi)$ and DDW gap
$D_\bk = {D_0 \over 2}(\cos k_x -\cos k_y )$. One obtains
\begin{equation}\label{eq:lam7}
\Lambda(\bq,i\omega) = {2\over L^2}{\sum_\bk}' 
{-\Omega_+\Omega_-+\epsilon''_+\epsilon''_-+D_+D_-\over
(\Omega_+^2+E_+^2)(\Omega_-^2+E_-^2)},
\end{equation}
with $i\Omega_\pm=i\omega-\epsilon'_\pm$,
$E_\pm=\sqrt{(\epsilon''_\pm)^2+D_\pm^2}$ and `$\pm$' denoting
$\bk\pm\bq/2$ as before.
At low energies the spectrum of a DDW quasiparticle is Dirac-like with 
the node fixed at $(\pi/2,\pi/2)$.  In this limit (\ref{eq:lam7}) 
can be again evaluated 
analytically by performing the nodal approximation. We find that
\begin{figure}
\includegraphics[width = 8cm]{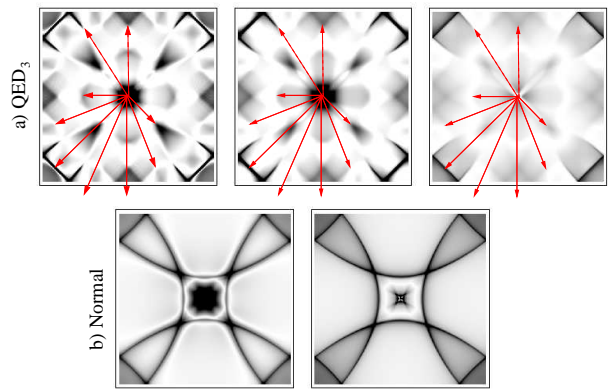}
\caption {a) Interference patterns within the QED$_3$ scenario 
with the anomalous dimension exponent  $\eta = 0.0, 0.2,0.4$ (left to right)
and all other parameters as in Fig.\ \ref{fig:dSC}.
b) The FT-STS pattern calculated for a normal metal with the dispersion
as in the $d$SC case and $\omega = 0.02t$ (left) 
and $\omega = 0.045t$ (right).}
\label{fig:eta}
\end{figure}
the result is given by Eq.\ (\ref{eq:lam5}) with $\omega$ replaced 
by $\omega+\mu$ and $v_\Delta$ by $v_D$, the slope of the DDW gap.
Thus, we find that at low energies DDW state will produce FT-STS
patterns similar to those expected for $d$SC with {\em magnetic}
scattering, characterized by continuous {\em lines} in the Brillouin
zone as opposed to the sharp peaks. This is confirmed by a full
numerical evaluation of Eq.\ (\ref{eq:lam7}) displayed
in Fig.\ \ref{fig:DDW}. The DDW patterns are markedly different 
from $d$SC even at half filling.

We now consider the class of theories which describe the pseudogap
as an incipient superconducting gap. In these theories the long range
superconducting order is destroyed above $T_c$ but short range 
pairing correlations persist for $T_c<T<T^*$. 
Based on the above discussion we expect that 
such a pseudogap phase would inherit the FT-STS patterns characteristic
of the superconducting phase since the coherence factors will 
locally retain their p-h character. As an example we
consider the QED$_3$ theory of the pseudogap phase which has been
proposed to describe a phase disordered $d$SC \cite{ft1}.
In this theory fluctuating phase of the SC order parameter 
produces an emergent massless U(1) gauge field which mediates
long range interactions between the fermions. As a result the fermion 
propagator becomes incoherent \cite{ftv1}: 
\begin{equation}\label{eq:luttinger}
G_0(\bk,i\omega)= {i\omega + \epsilon_\bk\tau_3 
\over [\omega^2+\epsilon_\bk^2+\Delta_\bk^2]^{1-\eta/ 2}},
\end{equation}
and exhibits a Luttinger liquid like 
dynamics at long distances with the sharp quasiparticle poles
replaced by branch cut singularities characterized by  
the anomalous dimension exponent $\eta$.
The latter is believed  to be a small positive
number but its exact value is a matter of debate \cite{ftv1}. 
Here we treat $\eta$ as a parameter and show that the structure
of FT-STS patterns is insensitive to its exact value.
The QED$_3$ propagator lacks the off-diagonal part, reflecting
the absence of true SC long range order.  

One can again evaluate Eq.\ (\ref{eq:lam1}) with the QED$_3$ propagator
(\ref{eq:luttinger}) analytically within the single node approximation. 
For scalar disorder one finds a result similar to Eq.\ (\ref{eq:lam4})
with the square root singularity at $E_\bq=2\omega$ replaced by a weaker 
$1/(z^2-1)^{{1\over 2}-\eta}$ singularity. Thus, for $\eta<{1\over 2}$
we expect patterns similar to those in $d$SC. 
The numerical results, presented in Fig.\ \ref{fig:eta}, indeed show 
that the QED$_3$ interference patterns retain peaks at the 
same positions as in the SC phase but the peaks become smeared 
for larger 
$\eta$ reflecting the incoherent nature of the fermionic excitations
described by Eq.\ (\ref{eq:luttinger}).

Finally, Fig.\ \ref{fig:eta} displays 
the FT-STS patterns for the normal
metallic state $(\Delta_0=D_0=0)$, which are expected to describe strongly 
overdoped cuprates. 

In conclusion, we have shown here that the quasiparticle interference 
patterns seen in the FT-STS reveal the signatures of both the 
quasiparticle dispersion and, through their sensitivity to
the quasiparticle coherence factors, the {\em nature of the underlying 
electronic order} present in the system. 
In particular, the latter determines the basic
characteristic features of the interference patterns. 
We have demonstrated, by general arguments and
detailed calculations within several relevant models, that the 
superconducting
order is very special in that it alone produces patterns
consistent with the experimental data. Based on this
insight we have proposed a test for the nature of the pseudogap phase
in cuprates using FT-STS. If the pseudogap is due to fluctuating SC 
order, then the FT-STS patterns above $T_c$ should remain 
qualitatively the same as those below $T_c$. If, on 
the other hand, the pseudogap is due to static or fluctuating order in the
p-h channel, such as SDW, CDW or DDW, the patterns above $T_c$ should
be {\em qualitatively different}, generally exhibiting 
continuous lines or peaks at different positions. 

As mentioned above preliminary experimental data
on Bi-2212 \cite{Yazdani} show FT-STS patterns above $T_c$ that are 
very  similar to those found  deep in the SC phase
\cite{Hoffman,McElroy,Howald}, suggesting that the pseudogap state
is of predominantly superconducting origin. One may ask how 
robust is this identification of the pseudogap state, provided that 
the experimental data  \cite{Yazdani} can be reproduced. 
Our analysis indicates that the p-p nature of superconducting
correlations plays critical role in formation of the patterns 
observed experimentally. Furthermore, despite significant effort, we 
were unable to construct a model with instability in p-h channel
that would mimic these patterns. Therefore, we must conclude that
 data of Yazdani and co-workers \cite{Yazdani}, if correct, place 
very strong constraints on the nature of the pseudogap state in
cuprates.

The authors are indebted to J.C. Davis, T. Davis, J. Hoffman, D.-H. Lee, 
S. Sachdev, D.E. Sheehy, O. Vafek, Z. Te\v{s}anovi\'c and A. Yazdani
for discussions and correspondence. This work was supported by NSERC,
CIAR and the A.P. Sloan Foundation.

\end{document}